\documentclass[12pt,a4paper]{article}
\usepackage{latexsym,amssymb,amsmath}

\newtheorem{theorem}{Theorem}
\newtheorem{lemma}{Lemma}

\newcommand{\rk}{{\rm rk}\,}

\newcommand{\proof}{\noindent {\it Proof. }}

\begin{document}
\title{On additive complexity of a sequence of matrices
\footnote{Research supported in part by RFBR, grants
11--01--00508, 11--01--00792, and OMN RAS ``Algebraic and
combinatorial methods of mathematical cybernetics and information
systems of new generation'' program (project ``Problems of optimal
synthesis of control systems'').}}
\date{}
\author{Igor S. Sergeev\footnote{e-mail: isserg@gmail.com}}

\maketitle

\section{Introduction}

The present paper deals with the complexity of computation of a
sequence of Boolean matrices via universal commutative additive
circuits, i.e. circuits of binary additions over the group
$(\mathbb Z,\, +)$ (an additive circuit implementing a matrix over
$(\mathbb Z,\, +)$, implements the same matrix over any
commutative semigroup $(S,\, +)$.) Basic notions of circuit and
complexity see in~\cite{eju,elu}.

Denote the complexity of a matrix $A$ over $(\mathbb Z,\, +)$ as
$L(A)$. Consider a sequence of $n\times n$-matrices $A_n$ with
zeros on the leading diagonal and ones in other positions. It is
known that $L(A_n)=3n-6$, see e.g.~\cite{ech}.

In~\cite{efin} it was proposed a sequence of matrices $B_{p,q,n}$
more general than $A_n$ and the question of complexity of the
sequence was investigated. Matrix $B_{p,q,n}$ has $C_n^q$ rows and
$C_n^p$ columns. Rows are indexed by $q$-element subsets of
$[1..n]$; columns are indexed by $p$-element subsets of $[1..n]$
(here $[k..l]$ stands for $\{k, k+1,\ldots, l\}$). A matrix entry
at the intersection of $Q$-th row and $P$-th column is 1 if $Q\cap
P = \emptyset$ and 0 otherwise.

Consider some simple examples of $B_{p,q,n}$. If $n<p+q$ then
$B_{p,q,n}$ is zero matrix. Evidently, $B_{1,1,n}=A_n$. By the
symmetry of definition $B_{p,q,n}=B^T_{q,p,n}$. Matrices
$B_{p,0,n}$ and $B_{0,q,n}$ are all-ones row and column
respectively. So, $L(B_{p,0,n})=C_n^p-1$, $L(B_{0,q,n})=0$.

Note that by the transposition principle (see e.g.~\cite{eju})
complexity of matri\-ces $B_{p,q,n}$ and $B_{q,p,n}$ satisfies the
identity
$$ L(B_{q,p,n}) = L(B_{p,q,n}) + C_n^q-C_n^p. $$

It was shown in~\cite{efin} that $L(B_{p,q,n})=O((n^p+n^q)\log
n)$. We prove better bound
$$ L(B_{p,q,n})\le(\alpha^p-1)C_n^q + \alpha^qC_n^p,$$
where $\alpha=\frac{3+\sqrt5}2$. This bound is linear (and
consequently tight up to a constant factor) for a constant $p$ and
$q \le 0.65n$.

The following lower bound
$$ L(B_{p,q,n})\ge(q-p+1)\sum_{k=0}^pC_n^k - 2^{p+q},$$
valid for $1\le p\le q$ and $n>p+q$, shows that the complexity of
$B_{p,q,n}$ is generally non-linear. For instance, one can try $p$
and $q$ of type $\frac{n}2-\Theta(\sqrt n)$ to obtain
$L(B_{p,q,n})=\Omega(N\log N)$, where $N=C_n^p+C_n^q$.

\section{Algorithm}

Let us introduce some notation. Let $\langle p,q,S_0,S\rangle$
denote a set of sums $\displaystyle y_Q = \sum_{P\subset
S\setminus Q,\, |P|=p} x_{S_0 \cup P}$, where $Q \subset S$,
$|Q|=q$. Thus, $\langle p,q,\emptyset,[1..n]\rangle$ is a result
of multiplication of the matrix $B_{p,q,n}$ by the vector of
variables $x_P$, $P\subset[1..n]$, $|P|=p$.

Let $\langle p, q,\emptyset, [1..n-1] \rangle$ is already computed
(with complexity $L(B_{p,q,n-1})$). We are to compute $\langle p,
q,\emptyset, [1..n] \rangle$. The computation consists of three
parts.

{\bf 1. Computation of $y_Q$, $\{1,\,n\} \cap Q = \emptyset$.}

1.1. Connect each input $x_{\{1\}\cup S}$ of a circuit computing
$\langle p, q,\emptyset, [1..n-1] \rangle$ with the following
precomputed sum
$$ x_{\{1\}\cup S} + x_{\{n\}\cup S},  \quad \text{if $2 \notin S$},  $$
$$ \sum_{T \subset ([1..k] \cup \{n\}), \; |T|=k} x_{T \cup S'},  \quad \text{if $S = [2..k] \sqcup S'$ and $(k+1) \notin S$}, \quad k \le p-1. $$
Note that in the sums above each variable $x_{\{n\}\cup S}$ occurs
exactly once. Thus, these sums can be computed with complexity
$C_{n-1}^{p-1}$.

1.2. Consider functioning of outputs of the transformed circuit.
Take an output implementing a sum $y_Q \in \langle p, q,\emptyset,
[1..n-1] \rangle$ in the original circuit. If $1 \in Q$, then
functioning of the output remained intact after transformation
since $y_Q$ depends on inputs which haven't changed. If $[1..k]
\cap Q = \emptyset$ and $(k+1) \in Q$, $1\le k \le p-1$, then the
output in the transformed circuit computes a sum
\begin{equation}\label{e1}
\sum_{P \cap Q = \emptyset,\;
|P|=p,\; ([1..k]\cup\{n\}) \not\subset P} x_P.
\end{equation}
To obtain a sum $y_Q \in \langle p, q,\emptyset, [1..n] \rangle$
one has to add summands $x_P$, $([1..k]\cup\{n\}) \subset P$, to
the sum~(\ref{e1}). At last, if $[1..p] \cap Q = \emptyset$, then
the output correctly computes a sum $y_Q \in \langle p,
q,\emptyset, [1..n] \rangle$ in the transformed circuit.

1.3. For any $k \in [1..p-1]$ compute
$$\langle p-k-1, q-1,[1..k]\cup\{n\}, [k+2..n-1] \rangle.$$
These are all sums needed to complete sums~(\ref{e1}) to obtain
$\langle p, q,\emptyset, [1..n] \rangle$.

The complexity of the computations can be estimated as
$$  \sum_{k=2}^{p} L(B_{p-k,q-1,n-k-1}). $$

1.4. Add the sums computed on the step 1.3 to sums~(\ref{e1}).
Complexity of this addition is the number of sums~(\ref{e1}), i.e.
the number of $q$-element sets $Q \subset [2..n-1]$ such that
$[2..p] \cap Q \ne \emptyset$. The latter number is
$C_{n-2}^q-C_{n-p-1}^q$.

{\bf 2. Computation of $y_Q$, $|\{1,\,n\} \cap Q|=1$.}

2.1. In the current circuit consider outputs implementing sums
$y_Q \in \langle p, q,\emptyset, [1..n-1] \rangle$, $1 \in Q$
(this outputs implemented the same sums in the original circuit).
Each such sum can be expanded to a sum $y_Q \in \langle p,
q,\emptyset, [1..n] \rangle$, $1 \notin Q$, $n \in Q$
(alternatively, $1 \in Q$, $n \notin Q$), via addition of summands
$x_{P}$, $1 \in P$, $P \subset [1..n-1]$ (respectively, $n \in P$,
$P \subset [2..n]$).

2.2. Compute sets $\langle p-1, q-1,1, [2..n-1] \rangle$ and
$\langle p-1, q-1,n, [2..n-1] \rangle$ with complexity
$2L(B_{p-1,q-1,n-2})$.

2.3. Add the last computed sums to the sums $y_Q \in \langle p,
q,\emptyset, [1..n-1] \rangle$, $1 \in Q$. It requires
$2C_{n-2}^{q-1}$ elementary additions.

{\bf 3. Computation of $y_Q$, $\{1,\,n\} \subset Q$.}

3.1. Note that any $q$-element set $Q \subset[1..n]$, $\{1,\,n\}
\subset Q$, satisfies condition: $[1..k-1] \subset Q$, $n \in Q$,
$k \notin Q$ for some $k \in [2..q]$.

Let $k \in [2..q]$. In the current circuit consider outputs
implementing sums $y_Q \in \langle p, q,\emptyset, [1..n-1]
\rangle$, $[1..k] \subset Q$, $(k+1) \notin Q$. (This set can be
defined alternatively as $\langle p, q-k,\emptyset, [k+1..n-1]
\rangle$.) Such sum can be expanded to a sum $y_Q \in \langle p,
q,\emptyset, [1..n] \rangle$, $[1..k-1] \subset Q$, $n \in Q$, $k
\notin Q$, via addition of appropriate summands $x_P$, $k \in P$,
$P \subset [k..n-1]$. The supplementing sums constitute the set
$\langle p-1, q-k,k, [k+1..n-1] \rangle$.

3.2. For any $k \in [2..q]$ compute the set $\langle p-1, q-k,k,
[k+1..n-1] \rangle$. It requires complexity
$$ \sum_{k=2}^q L(B_{p-1,q-k,n-k-1}). $$

3.3. Add the latter computed sums to the sums $y_Q \in \langle p,
q,\emptyset, [1..n-1] \rangle$ according to the item 3.1. It
requires $C_{n-2}^{q-2}$ elementary additions, by the number of
results.

\section{Upper bound}

The argument of the previous section leads to inequality:
\begin{multline}\label{e2}
L(B_{p,q,n}) \le L(B_{p,q,n-1}) + C_{n-1}^{p-1}+C_n^q-C_{n-p-1}^q
+ \\ + \sum_{k=1}^p L(B_{p-k,q-1,n-k-1}) + \sum_{k=1}^q
L(B_{p-1,q-k,n-k-1}),
\end{multline}
due to identity $C_{n-2}^q+2C_{n-2}^{q-1}+C_{n-2}^{q-2} = C_n^q$.

\begin{theorem}
Let $\alpha=\frac{3+\sqrt5}2$. Then
$$ L(B_{p,q,n}) \le (\alpha^p-1)C_n^q + \alpha^qC_n^p. $$
\end{theorem}

\proof The statement of the theorem is evidently holds when
$n=p+q$, or $p=0$, or $q=0$ (see introduction). Let us assume the
validity of the statement for all triples of parameters
$p',q',n'$, where $p'\le p$, $q'\le q$, $n'<n$ and consider the
triple $p,q,n$.

Put the assumed upper bounds in the second member of~(\ref{e2}).
To make calculations easier use identities:
$$ C_n^q-C_{n-p-1}^q = C_{n-1}^{q-1}+C_{n-2}^{q-1}+\ldots+C_{n-p-1}^{q-1} \le (p+1)C_{n-1}^{q-1}, $$
$$ C_n^0 + C_{n+1}^1 + \ldots C_{n+k}^k = C_{n+k+1}^k. $$
The last identity allows to estimate sums in~(\ref{e2}) as
following:
\begin{multline*}
\sum_{k=1}^p L(B_{p-k,q-1,n-k-1}) \le \alpha^{q-1}\sum_{k=1}^p
C_{n-k-1}^{p-k} + C_{n-1}^{q-1} \left( \sum_{k=0}^{p-1} \alpha^k -
p \right) \le \\ \le \alpha^{q-1}C_{n-1}^{p-1} + \left(
\frac{\alpha^p}{\alpha-1} - p-1 \right)C_{n-1}^{q-1},
\end{multline*}
\begin{multline*} \sum_{k=1}^q L(B_{p-1,q-k,n-k-1}) \le
(\alpha^{p-1}-1)\sum_{k=1}^q C_{n-k-1}^{q-k} + C_{n-1}^{p-1}
\sum_{k=0}^{q-1} \alpha^k \le \\ \le (\alpha^{p-1}-1)C_{n-1}^{q-1}
+ \left( \frac{\alpha^q}{\alpha-1} - 1 \right)C_{n-1}^{p-1}.
\end{multline*}

Finally, taking into account $1+\frac{\alpha}{\alpha-1}=\alpha$,
the second member of~(\ref{e2}) is bounded by
$$
(\alpha^p-1)C_{n-1}^q + \alpha^qC_{n-1}^p +
(\alpha^p-1)C_{n-1}^{q-1} + \alpha^qC_{n-1}^{p-1} \le
(\alpha^p-1)C_n^q + \alpha^qC_n^p,
$$
q.e.d.

\section{Lower bound}

\begin{lemma}
If $n\ge p+q$, then matrix $B_{p,q,n}$ has full rank over $\mathbb
R$.
\end{lemma}

\proof By invariance of rank with respect to transposition it is
sufficient to consider case $p\le q$ (so, $C_n^p \le C_n^q$).

We are to show that the rows of $B_{p,q,n}$ generate the space
$\mathbb R^{C_n^p}$. To be precise, we will prove that any vector
$(0,\ldots,0,1,0\ldots,0)$ with 1 in position~$P$ can be
represented as a linear combination of rows of $B_{p,q,n}$.

Let $a_0, \ldots, a_p \in \mathbb R$. Consider such linear
combination of rows, in which $Q$-th row occurs with the
coefficient $a_{|P\cap Q|}$. Clearly, such combination produces a
vector with coordinate in position $P'$ depending only on $|P \cap
P'|$. Denote the value of this coordinate as $b_{|P \cap P'|}$.

1. We are going to prove that a vector $(b_0,\ldots,b_p)^T$ is the
product of a vector $(a_p,\ldots,a_0)^T$ and some constant upper
triangular matrix $H$ with no zeros on the leading diagonal.

1.1. Firstly, check that $b_i$ depends on $a_{p-i}$ (hence, the
leading diagonal of $H$ contains no zeros). Indeed, let $P'
\subset [1..n]$ and $|P \cap P'|=i$. Consider a row indexed by
$Q$, $Q \cap P = P \setminus P'$, $Q \cap P' = \emptyset$. Such
row exists in view of inequality $n\ge p+q$. The row has 1 in
position $P'$ and it occurs in the linear combination with the
coefficient $a_{p-i}$.

1.2. Analogous argument shows that $b_i$ does not depend on
$a_{p-j}$ if $j<i$ (hence, all entries in $H$ below leading
diagonal are zero). Indeed, for any $Q$, $|Q \cap P| = p-j$, one
immediately concludes that $|Q \cap P'| \ge i-j
>0$. So the $Q$-th row has zero in position $P'$.

2. Therefore, for any vector $\bar b \in \mathbb R^{p+1}$, in
particular for the vector $(0,\ldots,0,1)$ we are interested in,
there exists a vector $\bar a \in \mathbb R^{p+1}$ such that $\bar
b = H \bar a$. The vector $\bar a$ defines the required linear
combination, q.e.d.

\begin{lemma}
Let $p\ge1$, $q\ge1$, $n> p+q$. Then
$$  L(B_{p,q,n}) \ge  L(B_{p,q-1,n-1}) +  L(B_{p-1,q,n-1}) + C_{n-1}^{\min\{p,\,q\}}. $$
\end{lemma}

\proof The proof of the lemma is similar to the proof of Th. 4
in~\cite{ebf}. Consider an arbitrary additive circuit $\Psi$
implementing $B_{p,q,n}$. Write $X_0 = \{ x_P \mid  n \notin P\}$,
$X_1 = \{ x_P \mid  n \in P\}$.

1. Consider the subcircuit of $\Psi$ which does not depend on
inputs $X_0$. Particularly, it implements the set $\langle
p,q-1,\emptyset,[1..n-1]\rangle$ and consequently contains at
least $L(B_{p,q-1,n-1})$ gates.

2. Calculate the number of gates in $\Psi$ with both inputs
depending on inputs from $X_1$. These gates together form a
circuit derived from $\Psi$ by replacement of inputs from $X_0$ by
zeros. In particular, this circuit computes $\langle
p-1,q,n,[1..n-1]\rangle$. Thus, the number of gates in question is
at least $L(B_{p-1,q,n-1})$.

3. Now, consider the gates of $\Psi$ with one input depending on
$X_1$ and another input not depending on $X_1$. Denote as $Y$ a
set of sums of variables in $X_0$ implemented by non-depending on
$X_1$ inputs of the gates. Note that $|Y|$ is a lower bound for
the number of the considered gates. It can be also seen that $Y$
generates the set $\langle p,q,\emptyset,[1..n-1]\rangle$
containing $X_0$-parts of sums implementing by $\Psi$ and
depending on $X_1$. Thus, $|Y| \ge \rk B_{p,q,n-1}$.  As follows
from Lemma 1, $\rk B_{p,q,n-1} = C_{n-1}^{\min\{p,\,q\}}$.

By putting estimates of items 1--3 together one obtains the
required inequality.

\begin{theorem}
Let $n> p+q$ and $p \le q$. Then
$$  L(B_{p,q,n})\ge(q-p+1)\sum_{k=0}^pC_n^k - 2^{p+q}. $$
\end{theorem}

\proof The proof is by induction as in Th. 1. Put the cases $p=0$
and $p=q=1$ as a base of induction ($L(B_{1,1,n})\ge n-3$
evidently holds, see introduction).

1. If $p<q$ then by the Lemma 2 and induction hypothesis one has
\begin{multline*}
L(B_{p,q,n})\ge C_{n-1}^p + (q-p)\sum_{k=0}^p C_{n-1}^k +
(q-p+2)\sum_{k=0}^{p-1} C_{n-1}^k - 2^{p+q} = \\ =
(q-p+1)\sum_{k=1}^p (C_{n-1}^k + C_{n-1}^{k-1}) + (q-p+1) -
2^{p+q} = (q-p+1)\sum_{k=0}^pC_n^k - 2^{p+q}.
\end{multline*}

2. In the case $p=q$ use transposition property
$$L(B_{p,p-1,n}) = L(B_{p-1,p,n})+C_{n-1}^p-C_{n-1}^{p-1},$$
to obtain
\begin{multline*}
L(B_{p,p,n})\ge 2C_{n-1}^p- C_{n-1}^{p-1} +
4\sum_{k=0}^{p-1}C_{n-1}^k - 2^{2p} > \\ > C_{n-1}^p + 2
\sum_{k=0}^{p-1}C_{n-1}^k - 2^{2p} = \sum_{k=0}^pC_n^k - 2^{2p}.
\end{multline*}
It completes the proof.

{\bf Remark.} In fact, Lemma 2 allows to deduce slightly stronger
inequality
$$  L(B_{p,q,n}) \ge C_n^p + \sum_{k=0}^p (p+q-2k+1)C_n^k-2^{p+q+1}. $$

\end{document}


\title{Об аддитивной сложности одной последовательности матриц
\footnote{Работа выполнена при финансовой поддержке РФФИ, проекты
11--01--00508 и 11--01--00792, и программы фундаментальных
исследований ОМН РАН <<Алгебраические и комбинаторные методы
математической кибернетики и информационные системы нового
поколения>> (проект <<Задачи оптимального синтеза управляющих
систем>>).}}
\date{}
\author{И.С. Сергеев\footnote{эл. адрес: isserg@gmail.com}}

\maketitle

\section{Введение}

В настоящей работе рассматривается сложность вычисления одной
последовательности булевых матриц универсальными коммутативными
аддитивными схемами, т.е. схемами из функциональных элементов
сложения над группой $(\mathbb Z,\, +)$.\footnote{Аддитивная
схема, реализующая некоторую матрицу над $(\mathbb Z,\, +)$,
реализует ту же матрицу над любой коммутативной полугруппой $(S,\,
+)$.} Базовые понятия схем и сложности см. в~\cite{lu,ju}.

Сложность вычисления матрицы $A$ аддитивными схемами над $(\mathbb
Z,\, +)$ обозначим через $L(A)$. Рассмотрим последовательность
$n\times n$-матриц $A_n$ с нулями на главной диагонали и единицами
вне главной диагонали. Известно, что $L(A_n)=3n-62$, см.,
например,~\cite{ch}.

В работе~\cite{fin} предложена последовательность матриц
$B_{p,q,n}$, более общая, чем $A_n$, и поставлен вопрос о
сложности вычисления матриц последовательности. Матрица
$B_{p,q,n}$ имеет $C_n^q$ строк и $C_n^p$ столбцов: строки
нумеруются подмножествами множества $[1..n]$ мощности $q$, а
столбцы~--- подмножествами множества $[1..n]$ мощности $p$ (запись
$[k..l]$ обозначает $\{k, k+1,\ldots, l\}$); на пересечении строки
$Q$ и столбца $P$ стоит 1, если $Q\cap P = \emptyset$, и 0~--- в
противном случае.

Рассмотрим простейшие примеры матриц $B_{p,q,n}$. При $n<p+q$
матрица $B_{p,q,n}$ является нулевой. Матрица $B_{1,1,n}$
совпадает с $A_n$. В силу симметрии $B_{p,q,n}=B^T_{q,p,n}$.
Матрицы $B_{p,0,n}$ и $B_{0,q,n}$ являются соответственно строкой
и столбцом из всех единиц, поэтому $L(B_{p,0,n})=C_n^p-1$,
$L(B_{0,q,n})=0$.

Отметим, что согласно принципу транспонирования (см.~\cite{ju})
сложность матриц $B_{p,q,n}$ и $B_{q,p,n}$ подчиняется соотношению
$$ L(B_{q,p,n}) = L(B_{p,q,n}) + C_n^q-C_n^p. $$

В работе~\cite{fin} показано, что $L(B_{p,q,n})=O((n^p+n^q)\log
n)$. Мы доказываем верхнюю оценку
$$ L(B_{p,q,n})\le(\alpha^p-1)C_n^q + \alpha^qC_n^p,$$
где $\alpha=\frac{3+\sqrt5}2$. Эта оценка линейна (и, как
следствие, по порядку точна), например, при постоянном $p$ и $q
\le 0,65n$.

То, что при подходящем выборе $p$ и $q$ сложность матриц
$B_{p,q,n}$ может быть более чем линейной, доказывает нижняя
оценка
$$ L(B_{p,q,n})\ge(q-p+1)\sum_{k=0}^pC_n^k - 2^{p+q},$$
справедливая при $1\le p\le q$ и $n>p+q$. Подставляя $p$ и $q$
вида $\frac{n}2-\Theta(\sqrt n)$, получаем оценку
$L(B_{p,q,n})=\Omega(N\log N)$, где $N=C_n^p+C_n^q$.

\section{Алгоритм}

Введем некоторые обозначения. Пусть $\langle p,q,S_0,S\rangle$
обозначает множество линейных комбинаций $\displaystyle y_Q =
\sum_{P\subset S\setminus Q,\, |P|=p} x_{S_0 \cup P}$, где $Q
\subset S$, $|Q|=q$. Множество $\langle
p,q,\emptyset,[1..n]\rangle$ есть результат умножения матрицы
$B_{p,q,n}$ на вектор переменных $x_P$, $P\subset[1..n]$, $|P|=p$.

Пусть множество $\langle p, q,\emptyset, [1..n-1] \rangle$
вычислено (со сложностью \linebreak $L(B_{p,q,n-1})$). Покажем как
вычислить $\langle p, q,\emptyset, [1..n] \rangle$. Вычисление
распадается на три части.

{\bf 1. Вычисление $y_Q$, $\{1,\,n\} \cap Q = \emptyset$.}

1.1. На входы $x_{\{1\}\cup S}$ схемы, вычисляющей $\langle p,
q,\emptyset, [1..n-1] \rangle$, подадим:
$$ x_{\{1\}\cup S} + x_{\{n\}\cup S},  \quad \text{если $2 \notin S$},  $$
$$ \sum_{T \subset ([1..k] \cup \{n\}), \; |T|=k} x_{T \cup S'},  \quad \text{если $S = [2..k] \sqcup S'$ и $(k+1) \notin S$}, \quad k \le p-1. $$
Заметим, что в приведенных суммах каждая переменная $x_{\{n\}\cup
S}$ встречается ровно по одному разу, поэтому весь набор
вычисляется со сложностью $C_{n-1}^{p-1}$.

1.2. Рассмотрим функционирование выходов схемы после
преобразования. Пусть до преобразования некоторый выход
реализовывал сумму $y_Q \in \langle p, q,\emptyset, [1..n-1]
\rangle$. Если $1 \in Q$, то функционирование выхода не
изменилось, т.к. не изменялись входы, от которых зависит $y_Q$.
Если $[1..k] \cap Q = \emptyset$, но $(k+1) \in Q$, где $1\le k
\le p-1$, то в измененной схеме выход вычисляет сумму
$$\sum_{P \cap Q = \emptyset,\; |P|=p,\; ([1..k]\cup\{n\})
\not\subset P} x_P.  \eqno(1)$$
До суммы $y_Q \in \langle p, q,\emptyset, [1..n] \rangle$
сумме~(1) не хватает слагаемых $x_P$, $([1..k]\cup\{n\}) \subset
P$. Наконец, если $[1..p] \cap Q = \emptyset$, то в измененной
схеме рассматриваемый выход правильно вычисляет сумму $y_Q \in
\langle p, q,\emptyset, [1..n] \rangle$.

1.3. При любом $k \in [1..p-1]$ вычислим множество сумм
$$\langle p-k-1, q-1,[1..k]\cup\{n\}, [k+2..n-1] \rangle.$$
Тем самым мы вычислим все суммы, дополняющие суммы~(1) до сумм из
$\langle p, q,\emptyset, [1..n] \rangle$.

Сложность этих вычислений оценивается как
$$  \sum_{k=2}^{p} L(B_{p-k,q-1,n-k-1}). $$

1.4. Прибавим суммы, вычисленные в предыдущем пункте, к
суммам~(1). Сложность этого шага равна числу сумм~(1), т.е. числу
множеств $Q \subset [2..n-1]$ мощности $q$ таких, что $[2..p] \cap
Q \ne \emptyset$, а это число равно $C_{n-2}^q-C_{n-p-1}^q$.

{\bf 2. Вычисление $y_Q$, $|\{1,\,n\} \cap Q|=1$.}

2.1. В построенной схеме рассмотрим выходы, реализующие суммы $y_Q
\in \langle p, q,\emptyset, [1..n-1] \rangle$, $1 \in Q$ (эти
выходы реализовывали те же самые функции и до преобразований
схемы). Каждой такой сумме до суммы $y_Q \in \langle p,
q,\emptyset, [1..n] \rangle$, $1 \notin Q$, $n \in Q$, не хватает
слагаемых $x_{P}$, $1 \in P$, $P \subset [1..n-1]$, а до суммы
$y_Q$, $1 \in Q$, $n \notin Q$, ей не хватает слагаемых $x_{P}$,
$n \in P$, $P
\subset [2..n]$. 

2.2. Вычислим множества $\langle p-1, q-1,1, [2..n-1] \rangle$ и
$\langle p-1, q-1,n, [2..n-1] \rangle$ со сложностью
$2L(B_{p-1,q-1,n-2})$.

2.3. Сложим вычисленные суммы с суммами $y_Q \in \langle p,
q,\emptyset, [1..n-1] \rangle$, $1 \in Q$. Это требует
$2C_{n-2}^{q-1}$ операций.

{\bf 3. Вычисление $y_Q$, $\{1,\,n\} \subset Q$.}

3.1. Заметим, что любое $q$-элементное множество $Q
\subset[1..n]$, $\{1,\,n\} \subset Q$, удовлетворяет условию:
$[1..k-1] \subset Q$, $n \in Q$, $k \notin Q$ при некотором $k \in
[2..q]$.

Пусть $k \in [2..q]$. В построенной схеме рассмотрим выходы,
реализующие суммы $y_Q \in \langle p, q,\emptyset, [1..n-1]
\rangle$, $[1..k] \subset Q$, $(k+1) \notin Q$. (Это множество
сумм можно иначе записать как $\langle p, q-k,\emptyset,
[k+1..n-1] \rangle$.) Такую сумму можно дополнить до суммы $y_Q
\in \langle p, q,\emptyset, [1..n] \rangle$, $[1..k-1] \subset Q$,
$n \in Q$, $k \notin Q$, добавив в нее подходящие слагаемые $x_P$,
$k \in P$, $P \subset [k..n-1]$. Дополняющие суммы образуют
множество $\langle p-1, q-k,k, [k+1..n-1] \rangle$.

3.2. При любом $k \in [2..q]$ вычислим множество $\langle p-1,
q-k,k, [k+1..n-1] \rangle$. Это выполняется со сложностью
$$ \sum_{k=2}^q L(B_{p-1,q-k,n-k-1}). $$

3.3. Прибавим суммы из предыдущего пункта к подходящим суммам $y_Q
\in \langle p, q,\emptyset, [1..n-1] \rangle$ согласно п. 3.1. Для
этого требуется $C_{n-2}^{q-2}$ сложений, по числу результатов.

\section{Верхняя оценка}

\setcounter{equation}{1}

Из оценок предыдущего раздела вытекает соотношение:
$$
\begin{array}{rl}
L(B_{p,q,n}) \le L(B_{p,q,n-1}) + C_{n-1}^{p-1}+C_n^q-C_{n-p-1}^q
& \mkern-14mu + \\ \displaystyle  + \sum_{k=1}^p
L(B_{p-k,q-1,n-k-1}) + & \mkern-10mu \displaystyle \sum_{k=1}^q
L(B_{p-1,q-k,n-k-1}),
\end{array} \eqno(2)
$$
поскольку $C_{n-2}^q+2C_{n-2}^{q-1}+C_{n-2}^{q-2} = C_n^q$.

\begin{theorem}
Пусть $\alpha=\frac{3+\sqrt5}2$. Тогда
$$ L(B_{p,q,n}) \le (\alpha^p-1)C_n^q + \alpha^qC_n^p. $$
\end{theorem}

\proof Утверждение теоремы очевидно выполнено при $n=p+q$ и в
случаях $p=0$ или $q=0$ (см. введение). Докажем его для тройки
параметров $p,q,n$ в предположении, что для троек $p',q',n'$, где
$p'\le p$, $q'\le q$, $n'<n$, оно уже доказано.

Подставим в правую часть~(2) оценки из индуктивного предположения.
Для упрощения выкладок используем соотношения:
$$ C_n^q-C_{n-p-1}^q = C_{n-1}^{q-1}+C_{n-2}^{q-1}+\ldots+C_{n-p-1}^{q-1} \le (p+1)C_{n-1}^{q-1}, $$
$$ C_n^0 + C_{n+1}^1 + \ldots C_{n+k}^k = C_{n+k+1}^k. $$
Последнее соотношение позволяет оценить суммы в~(2) как
\begin{multline*}
\sum_{k=1}^p L(B_{p-k,q-1,n-k-1}) \le \alpha^{q-1}\sum_{k=1}^p
C_{n-k-1}^{p-k} + C_{n-1}^{q-1} \left( \sum_{k=0}^{p-1} \alpha^k -
p \right) \le \\ \le \alpha^{q-1}C_{n-1}^{p-1} + \left(
\frac{\alpha^p}{\alpha-1} - p-1 \right)C_{n-1}^{q-1},
\end{multline*}
\begin{multline*} \sum_{k=1}^q L(B_{p-1,q-k,n-k-1}) \le
(\alpha^{p-1}-1)\sum_{k=1}^q C_{n-k-1}^{q-k} + C_{n-1}^{p-1}
\sum_{k=0}^{q-1} \alpha^k \le \\ \le (\alpha^{p-1}-1)C_{n-1}^{q-1}
+ \left( \frac{\alpha^q}{\alpha-1} - 1 \right)C_{n-1}^{p-1}.
\end{multline*}

Окончательно, учитывая, что $1+\frac{\alpha}{\alpha-1}=\alpha$,
правая часть~(2) оценивается как
$$
(\alpha^p-1)C_{n-1}^q + \alpha^qC_{n-1}^p +
(\alpha^p-1)C_{n-1}^{q-1} + \alpha^qC_{n-1}^{p-1} \le
(\alpha^p-1)C_n^q + \alpha^qC_n^p,
$$
что и требовалось доказать.

\section{Нижняя оценка}

\begin{lemma}
Если $n\ge p+q$, то матрица $B_{p,q,n}$ имеет полный ранг над
$\mathbb R$.
\end{lemma}

\proof В силу инвариантности ранга относительно транспонирования
достаточно рассмотреть случай $p\le q$ (при этом $C_n^p \le
C_n^q$).

Докажем, что строки матрицы $B_{p,q,n}$ порождают пространство
$\mathbb R^{C_n^p}$. Для этого покажем, что произвольный вектор
$(0,\ldots,0,1,0\ldots,0)$ с единицей на позиции с номером $P$
может быть представлен как линейная комбинация строк матрицы
$B_{p,q,n}$.

Пусть $a_0, \ldots, a_p \in \mathbb R$. Рассмотрим линейную
комбинацию строк, в которую строка с номером $Q$ входит с
коэффициентом $a_{|P\cap Q|}$. Ясно, что указанная линейная
комбинация дает вектор, у которого координата с номером $P'$
зависит только от $|P \cap P'|$, обозначим ее значение через
$b_{|P \cap P'|}$.

1. Докажем, что вектор $(b_0,\ldots,b_p)^T$ получается из вектора
$(a_p,\ldots,a_0)^T$ умножением на некоторую постоянную
верхнетреугольную матрицу $H$, не имеющую нулей на главной
диагонали.

1.1. Проверим, что $b_i$ зависит от $a_{p-i}$ (следовательно,
элементы на главной диагонали матрицы $H$~--- ненулевые).
Действительно, пусть $P' \subset [1..n]$ и $|P \cap P'|=i$.
Рассмотрим строку с номером $Q$, $Q \cap P = P \setminus P'$, $Q
\cap P' = \emptyset$. Такая строка существует в силу $n\ge p+q$; в
ней на позиции $P'$ стоит единица, и эта строка входит в линейную
комбинацию с коэффициентом $a_{p-i}$.

1.2. Аналогичное рассуждение показывает, что $b_i$ не зависит от
$a_{p-j}$ при $j<i$ (следовательно, элементы ниже главной
диагонали матрицы $H$~--- нулевые). В этом случае при любом $Q$,
$|Q \cap P| = p-j$, неизбежно получаем, что $|Q \cap P'| \ge i-j
>0$. Поэтому на позиции $P'$ в строке с номером $Q$ стоит ноль.

2. Таким образом, для любого вектора $\bar b \in \mathbb R^{p+1}$,
в частности для интересующего нас вектора $(0,\ldots,0,1)$,
найдется вектор $\bar a \in \mathbb R^{p+1}$, такой, что $\bar b =
H \bar a$. Этот вектор $\bar a$ и определяет искомую линейную
комбинацию. Лемма доказана.

\begin{lemma}
Пусть $p\ge1$, $q\ge1$, $n> p+q$. Тогда
$$  L(B_{p,q,n}) \ge  L(B_{p,q-1,n-1}) +  L(B_{p-1,q,n-1}) + C_{n-1}^{\min\{p,\,q\}}. $$
\end{lemma}

\proof Доказательство этой леммы подобно доказательству теоремы 4
из~\cite{bf}. Рассмотрим произвольную аддитивную схему $\Psi$,
реализующую $B_{p,q,n}$. Обозначим $X_0 = \{ x_P \mid  n \notin
P\}$, $X_1 = \{ x_P \mid  n \in P\}$.

1. В этой схеме рассмотрим подсхему, не зависящую от входов $X_0$.
Эта подсхема реализует, в частности, множество $\langle
p,q-1,\emptyset,[1..n-1]\rangle$ и, следовательно, содержит не
менее $L(B_{p,q-1,n-1})$ элементов.

2. В схеме $\Psi$ подсчитаем число элементов, оба входа которых
зависят от переменных из множества $X_1$. Эти элементы составляют
схему, получаемую из исходной при подстановке нулей вместо всех
переменных из $X_0$. Поскольку эта схема вычисляет, в частности,
множество $\langle p-1,q,n,[1..n-1]\rangle$, то число
рассматриваемых элементов не меньше чем $L(B_{p-1,q,n-1})$.

3. Рассмотрим элементы схемы $\Psi$, один вход которых зависит от
$X_1$, а другой не зависит. Обозначим через $Y$ множество линейных
комбинаций переменных $X_0$ на не зависящих от $X_1$ входах
рассматриваемых элементов. Ясно, что $|Y|$ является нижней оценкой
числа рассматриваемых элементов. По построению множество $\langle
p,q,\emptyset,[1..n-1]\rangle$ зависящих от $X_0$ подсумм сумм,
реализуемых на выходах схемы, зависящих от $X_1$, состоит из
линейных комбинаций сумм $Y$. Следовательно $|Y| \ge \rk
B_{p,q,n-1}$.  Согласно лемме 1, $\rk B_{p,q,n-1} =
C_{n-1}^{\min\{p,\,q\}}$.

Объединяя оценки пунктов 1--3, получаем утверждение леммы.

\begin{theorem}
Пусть $n> p+q$ и $p \le q$. Тогда
$$  L(B_{p,q,n})\ge(q-p+1)\sum_{k=0}^pC_n^k - 2^{p+q}. $$
\end{theorem}

\proof Доказательство проведем индукцией как в теореме~1. В
основание индукции положим случаи $p=0$ и $p=q=1$ (неравенство
$L(B_{1,1,n})\ge n-3$ очевидно справедливо, см. введение).

1. В случае $p<q$ согласно лемме 2 и предположению индукции имеем
\begin{multline*}
L(B_{p,q,n})\ge C_{n-1}^p + (q-p)\sum_{k=0}^p C_{n-1}^k +
(q-p+2)\sum_{k=0}^{p-1} C_{n-1}^k - 2^{p+q} = \\ =
(q-p+1)\sum_{k=1}^p (C_{n-1}^k + C_{n-1}^{k-1}) + (q-p+1) -
2^{p+q} = (q-p+1)\sum_{k=0}^pC_n^k - 2^{p+q}.
\end{multline*}

2. В случае $p=q$, используя тождество принципа транспонирования
$$L(B_{p,p-1,n}) = L(B_{p-1,p,n})+C_{n-1}^p-C_{n-1}^{p-1},$$
получаем
\begin{multline*}
L(B_{p,p,n})\ge 2C_{n-1}^p- C_{n-1}^{p-1} +
4\sum_{k=0}^{p-1}C_{n-1}^k - 2^{2p} > \\ > C_{n-1}^p + 2
\sum_{k=0}^{p-1}C_{n-1}^k - 2^{2p} = \sum_{k=0}^pC_n^k - 2^{2p}.
\end{multline*}
Теорема доказана.

{\bf Замечание.} Более аккуратные рассуждения позволяют из
соотношения леммы 2 вывести оценку
$$  L(B_{p,q,n}) \ge C_n^p + \sum_{k=0}^p (p+q-2k+1)C_n^k-2^{p+q+1}. $$